\documentclass[aps, prb, preprint, amsmath,amssymb,showpacs,superscriptaddress, longbibliography]{revtex4-1}
\usepackage{graphicx}
\usepackage{color}
\definecolor{green}{RGB}{34,139,34}
\makeatletter

\begin{document}%

\title{Discovery of a weak topological insulating state and van Hove singularity in triclinic RhBi$_{2}$}

\author{Kyungchan Lee}
\affiliation{Ames Laboratory, Ames, Iowa 50011, USA}
\affiliation{Department of Physics and Astronomy, Iowa State University, Ames, Iowa 50011, USA}

\author{Gunnar F. Lange}
\affiliation{TCM Group, Cavendish Laboratory, University of Cambridge, Cambridge CB3 0HE, United Kingdom}

\author{Lin-Lin Wang}
\affiliation{Ames Laboratory, Ames, Iowa 50011, USA}
\affiliation{Department of Physics and Astronomy, Iowa State University, Ames, Iowa 50011, USA}

\author{Brinda Kuthanazhi}
\affiliation{Ames Laboratory, Ames, Iowa 50011, USA}
\affiliation{Department of Physics and Astronomy, Iowa State University, Ames, Iowa 50011, USA}

\author{Tha\'{i}s V. Trevisan}
\affiliation{Ames Laboratory, Ames, Iowa 50011, USA}
\affiliation{Department of Physics and Astronomy, Iowa State University, Ames, Iowa 50011, USA}

\author{Na Hyun Jo}
\affiliation{Ames Laboratory, Ames, Iowa 50011, USA}
\affiliation{Department of Physics and Astronomy, Iowa State University, Ames, Iowa 50011, USA}

\author{Benjamin Schrunk}
\affiliation{Ames Laboratory, Ames, Iowa 50011, USA}
\affiliation{Department of Physics and Astronomy, Iowa State University, Ames, Iowa 50011, USA}

\author{Peter P. Orth}
\affiliation{Ames Laboratory, Ames, Iowa 50011, USA}
\affiliation{Department of Physics and Astronomy, Iowa State University, Ames, Iowa 50011, USA}

\author{Robert-Jan Slager}
\email[]{rjs269@cam.ac.uk}
\affiliation{TCM Group, Cavendish Laboratory, University of Cambridge, Cambridge CB3 0HE, United Kingdom}
\affiliation{Department  of  Physics,  Harvard  University,  Cambridge  MA  02138,  USA}

\author{Paul C. Canfield}
\email[]{canfield@ ameslab.gov}
\affiliation{Ames Laboratory, Ames, Iowa 50011, USA}
\affiliation{Department of Physics and Astronomy, Iowa State University, Ames, Iowa 50011, USA}

\author{Adam Kaminski}
\email[]{adamkam@ameslab.gov}
\affiliation{Ames Laboratory, Ames, Iowa 50011, USA}
\affiliation{Department of Physics and Astronomy, Iowa State University, Ames, Iowa 50011, USA}

\date{\today}
\maketitle 
{\bf  Time reversal symmetric (TRS) invariant topological insulators (TIs) fullfil a paradigmatic role in the field of topological materials, standing at the origin of its development. 
Apart from TRS protected 'strong' TIs, it was realized early on that more confounding weak topological insulators (WTI) exist. WTIs depend on translational symmetry and exhibit topological surface states only in certain directions making it significantly more difficult to match the experimental success of strong TIs. We here report on the discovery of a WTI state in RhBi$_{2}$ that belongs to the optimal space group P$\bar{1}$, which is the only space group where symmetry indicated eigenvalues enumerate all possible invariants due to absence of additional constraining crystalline symmetries. Our ARPES, DFT calculations, and effective model reveal topological surface states with saddle points that are located in the vicinity of a Dirac point resulting in a van Hove singularity (VHS) along the (100) direction close to the Fermi energy ($E_{\textrm{F}}$). Due to the combination of exotic features, this material offers great potential as a material platform for novel quantum effects.}

From the perspective of considering solely TRS, 3D topological insulators are classified by a series of $\textbf{Z}_{2}$ topological invariants,~($\nu_{0}$;$\nu_{1}$$\nu_{2}$$\nu_{3}$). The case $\nu_{0}$\,=\,1 is referred to as a strong topological insulator (STI)~ \cite{hasan2010colloquium, RevModPhys.83.1057}. A characteristic feature of a STI is the existence of gapless topological surface state (TSS) on all  surfaces, which have been observed experimentally in many materials such as Bi$_{1-x}$Sb$_{x}$, Bi$_{2}$Se$_{3}$, and Bi$_{2}$Te$_{3}$\,\cite{lv2019angle}. 
In contrast to 2D TIs, even when $\nu_{0}$\,=\,0 the system can have non-trivial $\nu_i$, resulting in a weak topological insulator (WTI). Reported WTI states are rather rare as gapless TSS can be detected only on particular surfaces, which may not be natural cleaving planes in many 3D materials. For instance, Bi$_{14}$Rh$_{3}$I$_{9}$ was predicted to be a WTI with $\textbf{Z}_{2}$\,=\,(0;001). However, experimental verification of the TSS remain elusive since the cleaving plane is (001), which is a topologically trivial side\,\cite{rasche2013stacked_WTI}. The recent development of nano-ARPES has reinvigorated  the search for WTIs. For instance, a first WTI TSS was only recently measured by isolating a $\sim$\,2\,$\mu$m size side facet and using nano-ARPES in quasi-one-dimensional bismuth iodide $\beta$-Bi$_{4}$I$_{4}$ \cite{WTI_Bi4I4noguchi2019weak}.

As part of an effort to design and discover materials with novel, intrinsically non-trivial, topological properties we identified RhBi$_{2}$ as an exciting candidate material.  To minimize potential degeneracies in reciprocal space, we decided to start with the lowest possible symmetry unit cell: triclinic. To maximize spin-orbit coupling we chose the heaviest (essentially) stable element, Bi as a majority component. For simplicity we chose to start by limiting our search to binary compounds. As a result of this cascade of physically inspired constraints we rapidly identified RhBi$_{2}$ as a promising candidate material with clear van der Waals-like bonding along the crystallographic $a$-axis (Fig. 1b). Here, we provide theoretical understanding and experimental evidence of the simplest WTI state in triclinic crystal RhBi$_{2}$ with clear TSS at the natural cleaving surface. As we will discuss below, the WTI classification of RhBi$_{2}$ maximally profits from its space group. As $P\bar{1}$ only features inversion in addition to translations, all topological properties in the presence of TRS can faithfully be determined from the parity eigenvalues of the occupied bands at TRIM points \cite{fu2007TI}. Moreover, due to the absence of additional symmetries none of these high symmetry momenta are related and a WTI can exist in every stacking direction~\cite{Slager_G}. In fact, this carefully selected, simple structure makes this the only space group in which the symmetry eigenvalues at TRIM points convey all topological invariants one-to-one \cite{Song_SI,Kruthoff_combinatorics,TQC,symmetryindicators}. Rather surprisingly, we find that this freedom of choosing the stacking direction culminates in an index, (0;001), that is perpendicular to the layering of the material, creating the possibility of a TSS preserving cleavage plane.

The lack of symmetries, especially the absence of rotational symmetry, in triclinic RhBi$_{2}$  allows the system to have a saddle points, where the curvature of a band has opposite sign in two perpendicular direction in momentum space. A saddle point is closely related to a VHS, leading to a  divergence in the density of states. Previously, VHSs have been reported in a variety of materials; graphene \cite{2010twisted_graphene, 2010extendedVHS_graphene}, cuprate superconductors \cite{1990copper_oxide_SC,1993occurrence_VHS_SC,1995observation_VHS_SC}, Pt$_{2}$HgSe$_{3}$\,\cite{ghosh2019saddle}, Pb$_{1-x}$Sn$_{x}$Se(Te)\cite{neupane2015topological_saddle}.  The proximity of the $E_{\textrm{F}}$ to a VHS can significantly enhance quantum many body instabilities and it can drive superconductivity\cite{hirsch1986enhanced_VHS_SC,kohn1965_SC_VHS,honerkamp2001_VHS_SC} ferromagnetism \cite{gonzalez2000kinematics,ziletti2015_VHS_ferro,fleck1997ferror_VHS,hlubina1997ferromagnetism}and antiferromagnetism \cite{lederer1987antiferromagnetism} depending on the topology of the band structure. Owing to the simplest topological classification and VHS in the vicinity of $E_{\textrm{F}}$, triclinic crystal RhBi$_{2}$ offers unique advantages for exploring exotic quantum phenomena in a  weak topological insulator. 

We present a study of triclinic RhBi$_{2}$ by using ARPES, DFT calculations and an effective model. Our result reveals two surface Dirac points, which are protected by time-reversal symmetry at the $\bar{\Gamma}$  and $\bar{Z}$ points. We focused experimental  attention at the momentum space around the $\bar{Z}$ point of the surface BZ. The shape of the Dirac cone around $\bar{Z}$ is significantly modified from typical Dirac dispersion due to the presence of two saddle points associated with a VHS. The saddle points are located  at a binding energy of $\sim$80 meV. Based on DFT calculations, we identify that surface state of RhBi$_{2}$ along (100) has a WTI phase, which is characterized by \textbf{Z}$_{2}$=(0;001). 

\begin{figure*}[!ht]
    \includegraphics[scale=0.5] {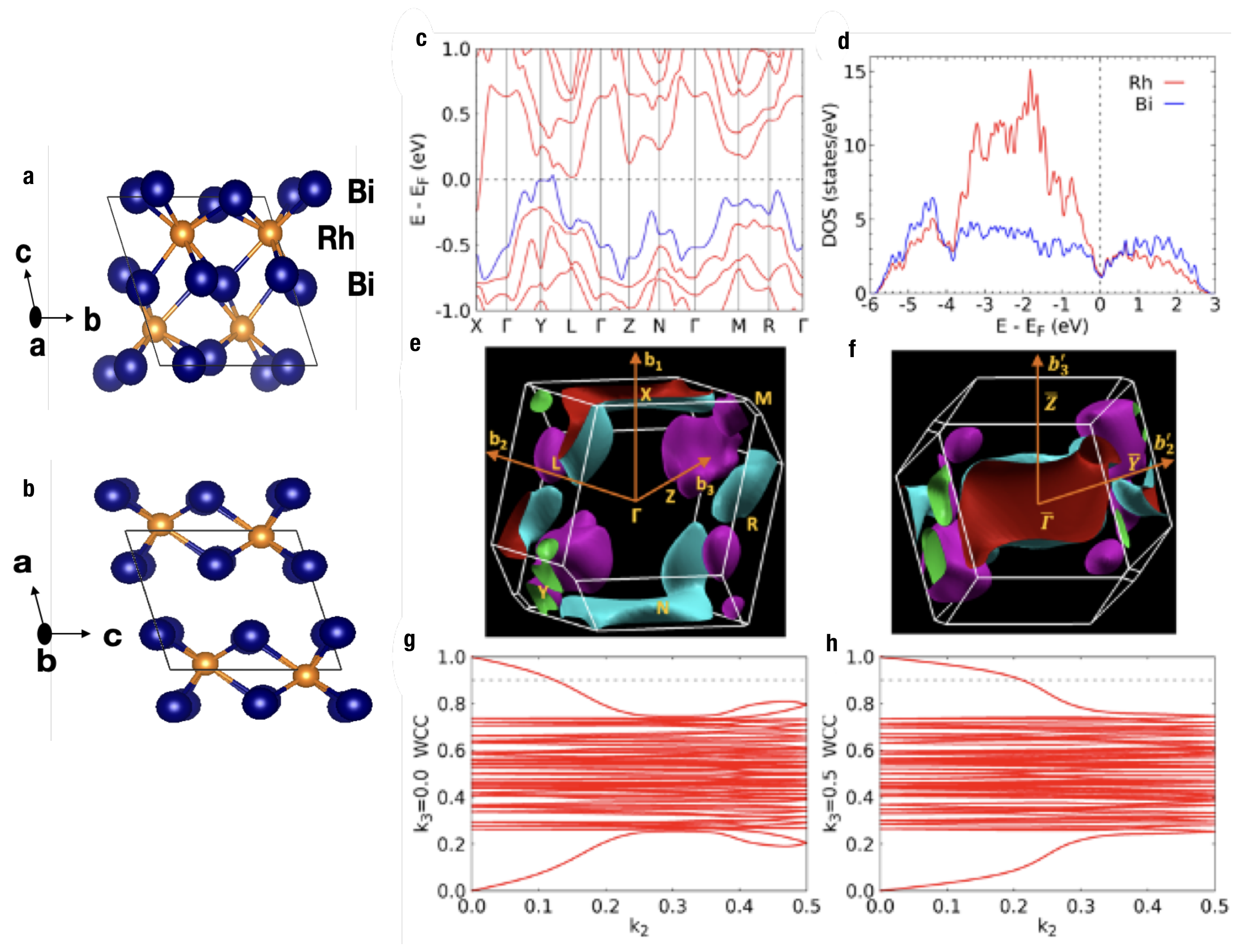}
    \caption{\textbf{Calculated bulk band structure, DOS and WCC evolution of RhBi$_{2}$.} \textbf{a}-\textbf{b}, Crystal structure of triclinic RhBi$_{2}$ with different orientations (Bi:blue spheres, Rh: yellow spheres). \textbf{c}, Bulk band structure of RhBi$_{2}$ in the triclinic crystal structure (space group 2) calculated in DFT-PBE with SOC. The top valence band according to electron filling is in blue. \textbf{d}, Density of states (DOS) projected on Rh and Bi.  \textbf{e}, 3D Fermi surface at the ($E_{\textrm{F}}$) with eight time-reversal invariant momentum (TRIM) points and reciprocal lattice vectors labeled. Electron and hole pockets are in cyan (red inside) and purple (green inside) respectively. \textbf{f}, 3D FS viewed along b$_{1}$ axis with projection of TRIM points and 2D reciprocal lattice vectors labeled. \textbf{g} and \textbf{h}, Wannier charge center (WCC) evolution on the k$_{3}$=0.0 and 0.5 planes, respectively,  showing the non-trivial topology.} 
 \label{fig:2}
\end{figure*}

Due to the symmetry group hosting only an inversion center, the layered triclinic RhBi$_{2}$ is an interesting case in terms of both band structure and topology. The bulk band structure of RhBi$_{2}$ calculated from density functional theory\cite{CAL_1_hohenberg1964inhomogeneous,CAL_2_kohn1965self} (DFT) using PBE exchange correlation functional with spin-orbit coupling (SOC) is plotted in Fig.\,\ref{fig:2}c. The 3D FS at the $E_{\textrm{F}}$ is shown in Fig.\,\ref{fig:2}e with the eight time-reversal invariant momentum (TRIM) points and reciprocal lattice vectors labeled. By virtue of inversion symmetry and TRS, all bands are doubly degenerated in the BZ due to Kramers theorem. The bands of interest arise by hybridization from Rh $d$ and Bi $p$ orbitals. From the small but finite density of states (DOS) at $E_{\textrm{F}}$ in Fig.\,\ref{fig:2}d, we see that RhBi$_{2}$ is a metal exhibiting non-overlapping pockets. There are no crossings between the top valence and bottom conduction bands anywhere in the BZ. The two bands crossing the $E_{\textrm{F}}$ give multiple electron and hole pockets. The electron and hole pockets are near BZ boundaries, while there is a large band gap around the BZ center $\Gamma$ point. Notable are the electron pockets around X and near L points and the hole pockets near the Y point. There are also two small hole pockets totally inside the BZ. Although the structure is stacked on the bc plane along the b$_{1}$ direction, a large electron pocket is on the b$_{1}$=0.5 plane around the X point. In contrast, there is band gap on the b$_{3}$=0.5 plane around the Z point.

With inversion being the only symmetry in the system besides time-reversal symmetry, the topological features are determined by the parity eigenvalues of the occupied bands at TRIM points \cite{fu2007TI}. In fact, from the perspective of the recent advancements in classifying band structures using constraints in symmetry eigenvalues \cite{Kruthoff_combinatorics} and comparing them to atomic configurations \cite{TQC, symmetryindicators}, the space group at hand, number 2 or $P\bar{1}$, is the {\it only} one in which such symmetry indicators unambiguously define all topological invariants \cite{Song_SI,Slager_G}, as there are no symmetry constraints between high symmetry momenta \cite{Slager_G}. The topological indices, apart from TRS invariant $\nu_0$ that we find to be zero, comprise three $\mathbf{Z}_2$ invariants corresponding to the weak indices and a $\mathbf{Z}_4$ index\cite{Khalaf_SI} that comes from promoting $\nu_0$ to the inversion symmetric context. 
The latter corresponds to the twice the value of the inversion invariant $\delta_i$, $\mathbf{Z}_4=2\delta_i$ , here\cite{Song_SI}. The calculated $\mathbf{Z}_4$ index is 2, indicating that RhBi$_{2}$ is topologically non-trivial, but indeed not a strong topological insulator. The calculated parity eigenvalue products are (+1) at $\Gamma$  and Z points, and (-1) at all the other TRIMs, which gives the Fu-Kane index of (0;001), determining the three $\mathbf{Z}_2$ invariants. This can also be confirmed by Wannier charge center (WCC) evolution or Wilson loop on k$_{3}$=0.0 and $0.5$ planes in Figs.\,\ref{fig:2}g and g, showing 2D $\mathbf{Z_{2}}$=1 on these planes \cite{Yu_Wilson,Aris_Wilson, Bouhon_Wilson}. In contrast, 2D $\mathbf{Z_{2}}$=0 on the other k$_{1,2}$=0.0 or 0.5 planes. Interestingly, although the layered structure is stacked along the b$_{1}$ direction, the weak TI can be seen as a stacking of quantum spin Hall (QSH) layers along the b$_{3}$ direction. 

In Fig.\,\ref{fig:2}f, the 3D FS is rotated for the view along b$_{1}$ or perpendicular to the \textit{bc} plane, i.e. (100). On the (100) surface, the $\Gamma$ and X points are projected to $\bar{\Gamma}$   point, Z and N to $\bar{Z}$ point, and Y and L to $\bar{Y}$  point, respectively. This shows that for the 2D FS on (100) around the $E_{\textrm{F}}$, the area near the $\bar{Z}$ point is gapped without bulk band projection, while the $\bar{\Gamma}$  and $\bar{Y}$ points are surrounded by bulk states. For bulk-boundary correspondence of TIs~\cite{BBC_2015}, parity eigenvalue products can also be projected onto surface as in $\pi_{a}$=$\delta_{a1}$ $\delta_{a2}$, where TRIM a$_{1}$ and a$_{2}$ projected into a on surface. The result is (-1) for $\bar{\Gamma}$   and  $\bar{Z}$ points, while (+1) for the others. Thus, an even number of two surface Dirac points will emerge on (100), one at $\bar{\Gamma}$   and the other at $\bar{Z}$. In summary, RhBi$_{2}$ is a topologically non-trivial 2D material as a layered compensating semimetal hosting the simplest WTI possible as the underlying space group has minimal symmetry.

\begin{figure*}[!htb]
    \includegraphics[scale=0.65] {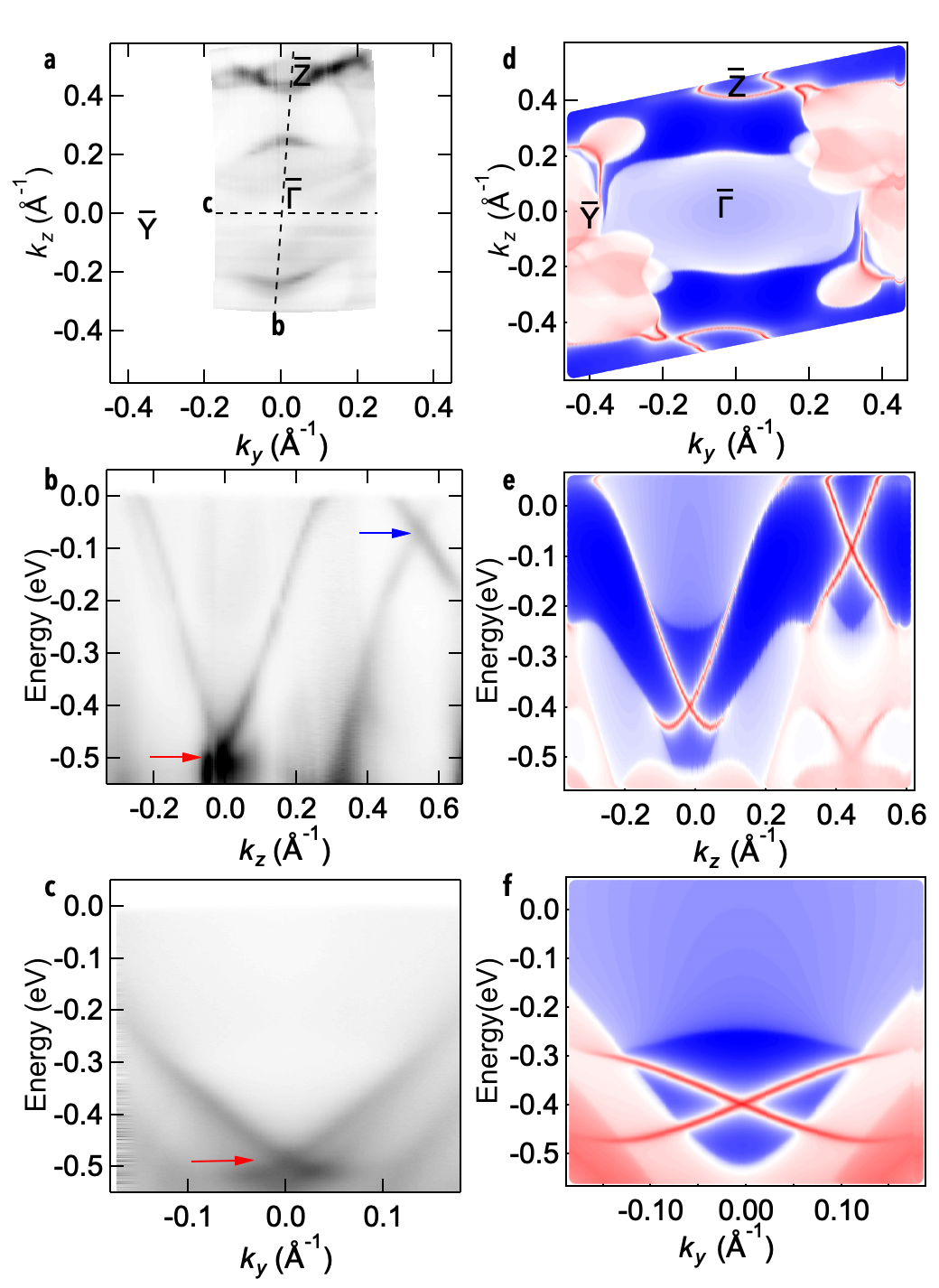}
    \caption{\textbf{Experimental and calculated Fermi surface and band dispersion of RhBi$_{2}$ at $T$~=~40~K.} \textbf{a}, Fermi surface plot of the ARPES intensity integrated within 10 meV of the chemical potential along $\bar{\Gamma}$-$\bar{Z}$.  \textbf{b}, Band dispersion along the $\bar{\Gamma}$-$\bar{Z}$ line. \textbf{c}, Band dispersion along $\bar{\Gamma}$-$\bar{Y}$. \textbf{d}, DFT calculations of Fermi surface. \textbf{e-f}, DFT calculations of energy dispersion corresponding to the ARPES data \textbf{b} and \textbf{c}, respectively.} 
 \label{fig:3}
\end{figure*}

In order to verify the nontrivial topological nature of triclinic RhBi$_{2}$, we performed ARPES measurements and DFT calculations to identify relevant features in the electronic structure. Fig.\,\ref{fig:3}a shows the FS of triclinic RhBi$_{2}$ measured using laser-based APRES, where dashed lines mark main crystallographic directions in the BZ. The measured band dispersion along those lines are shown in Figs.\,\ref{fig:3}b and c. Two Dirac points are clearly present in the data: one located at the $\bar{\Gamma}$ point with a binding energy of $\sim$\,500\,meV  and the other at the $\bar{Z}$ point with a binding energy of $\sim$\,80\,meV. The red arrow marks the Dirac point at the $\bar{\Gamma}$ point and the blue arrow marks the Dirac point at the $\bar{Z}$ point. Fig.\,\ref{fig:3}c shows the band dispersion along the $\bar{\Gamma}$-$\bar{Y}$ direction where the red arrow marks the location of the Dirac point there. Figs.\,\ref{fig:3}d-f show DFT calculations along the same directions as for the ARPES data. Overall, DFT calculations agree quite well with experimental data. Unlike typical TSS of STIs, which have a single Dirac point, we observe two surface Dirac points at TRIM points. This apparent discrepancy  between STIs and WTIs shows that triclinic RhBi$_{2}$ possesses TSS arising from weak topological insulating phase in (100) direction. In addition, we observed twofold rotational symmetry at the FS from the DFT calculations, although the crystal structure is triclinic. This is because three-dimensional inversion symmetry reduces to effectively two-dimensional rotation symmetry upon projection of all transverse momenta. To understand this phenomenon, we carefully studied different projections of each layer in the unit cell. Appendix C shows different projections of $k_{1}(k_{x})$ values on (100) surface. Red looped curves represent electron pockets. Based on the calculation, only projections from $k_{1}$= 0 and 0.5 planes show twofold rotational symmetry and the $k_{1}$ = 0 plane shows large electron pocket around the center of the BZ. Therefore, the large electron pocket around $\bar{\Gamma}$ is mainly from the $k_{1}$ = 0 plane and addition of all planes in the BZ will give us the effective twofold rotational symmetry on the 2D FS.    

\begin{figure}[!htb]
    \includegraphics[scale=0.65]{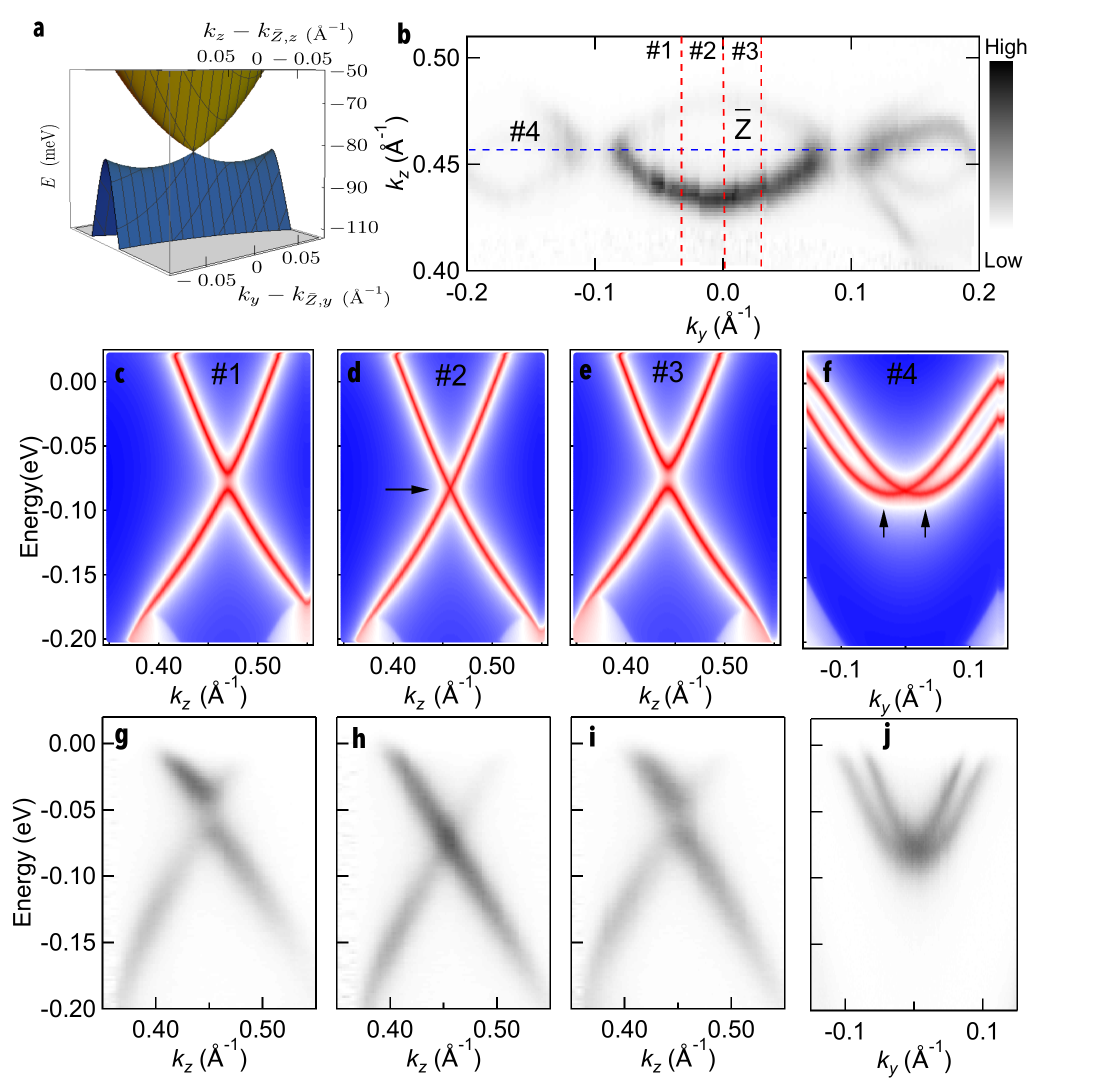}
    \caption{\textbf{Saddle points and Dirac dispersion around the $\bar{Z}$ point.} \textbf{a}, $k\cdot p$ model band dispersion around the $\bar{Z}$ point, revealing a nodal point, two saddle points and the almost-flat region of the lower energy band at slightly lower energies. The two saddle points at energy $-83$ meV are responsible for the logarithmic divergence of the DOS .\textbf{b}, Fermi surface plot of the ARPES intensity integrated within 10 meV of the chemical potential around the $\bar{Z}$ point. Dark areas mark location of the FS. 
    \textbf{c}-\textbf{f}, Calculated band dispersion along the vertical direction in \textbf{b} that is marked as red dashed lines (\#1-3), respectively.  \textbf{f}, Calculated band dispersion along the horizontal direction in \textbf{b} that is marked as a blue dashed line (\#4). Black arrows mark locations of band minimums.
    \textbf{g}-\textbf{j} ARPES data along the same cuts as in \textbf{c-f}.} 
 \label{fig:4}
\end{figure}

Out of the two surface Dirac points present, the Dirac point around the $\bar{Z}$ point is by far the more interesting. Instead of a typical Dirac like dispersion\,\cite{xia2009observationBi2Se3}, it shows a strong warping effect along the energy axis making saddle points in the lower band. $k\cdot p$ model band dispersion in Fig.\,\ref{fig:4}a illustrates the Dirac point and two saddle points that exist in proximity of the $\bar{Z}$ point. The FS in this area of BZ is plotted in Fig.\,\ref{fig:4}b. 
Figs.\,\ref{fig:4}c-f display calculated DFT band dispersion in the vertical direction(\#1-3) and one horizontal cut($\#$4) which are marked as dashed lines in Fig.\,\ref{fig:4}b. Interestingly, the vertical cuts (along $k_{z}$), Figs.\,\ref{fig:4}c-e, show the lower band with negative curvature while the horizontal cut (along $k_{y}$), Fig.\,\ref{fig:4}f, shows the lower  band with positive curvature. Along cut 4, there are actually two upper bands that give rise to two local minimums in energy as marked by black arrows. Those two minimums give rise to two saddle points as demonstrated in Fig.\,\ref{fig:4}a. In cut 1 we can see the upper and lower band are separated in energy (Fig.\,\ref{fig:4}c). Band dispersion along  the $\bar{Z}$-$\bar{\Gamma}$ line is shown in Fig.\,\ref{fig:4}d. Dirac dispersion is clearly visible with Dirac point at the binding energy of $\sim$\,80~meV and is marked by the black arrow. Figure\,\ref{fig:4}e displays the band dispersion at the cut 3, where the two bands are clearly separated as in cut 1. The separation in Figs.\,\ref{fig:4}c  and e is fairly symmetric with respect to the Dirac point. Corresponding ARPES data, measured along the same cuts in the BZ are shown in Figs.\,\ref{fig:4}g-j. Based on calculations and the experimental result, we conclude that this is the surface state forms a Dirac point rather than a line of Dirac dispersion along the $k_{y}$ direction. On the other hand, band dispersion in the horizontal direction significantly deviates from the vertical directions as shown in Fig.\,\ref{fig:4}f and j. Both upper and lower bands have a parabolic dispersion instead of the linear dispersion. Naturally, the lower band shows positive curvature. As a result, it looks like the overlap of two parabolic bands with slightly different curvatures.  


To qualitatively address the observed saddle points in vicinity of  the $\overline{Z}$ point on the (100) surface we consider a simple effective model.
As we have time-reversal ($\cal{T}$) and inversion symmetry ($\cal{P}$) in the bulk, $\cal{A}=\cal{PT}$ commutes with Hamiltonian, reflecting the Kramers degeneracy by virtue of squaring to $-1$. The presence of this symmetry will constrain the form of the bulk Hamiltonian affecting the effective model for the surface after projecting.  This will be reported elsewhere and we simply note that $\cal{P}$ is formally broken on the surface. In fact, having the inversion centre at the origin ensures that surface states with opposite $\cal{P}$-value reside on opposite surfaces. Hence, as motivated in SI Appendix B, we consider a simple two-band model that is only constrained by TRS to describe the strongly anisotropic anomalous edge states,
\begin{equation}
    H(\mathbf{k}) = d_0(\mathbf{k})I + d_1(\mathbf{k})\sigma_x+d_2(\mathbf{k})\sigma_y  + d_3(\mathbf{k})\sigma_z
    \label{eq:h}
\end{equation}
In the above $d_0(\mathbf{k})$ is even in $\mathbf{k}$ and $d_{1,2,3}(\mathbf{k})$ are odd in $\mathbf{k}$. Expanding our energy to second order then gives
\begin{equation}
\varepsilon_{\pm}(\mathbf{k}) = E_0+A_1k_y^2 + A_2k_z^2 + A_3k_yk_z \pm \sqrt{\sum_{i=1}^3 (B_i k_y + C_ik_z)^2},
\label{eq:g}
\end{equation}
where $E_0$ is the energy at the crossing points of the surface bands, $\{A_i\},\{B_i\}$ and $\{C_i\}$ are all real parameters and momenta are measured relative to the $\overline{Z}$ point. To accommodate strain effects, we could also expand $H(\mathbf{k})$ to third order, giving:
\begin{equation}
\varepsilon_{3\pm}(\mathbf{k}) = E_0+A_1k_y^2 + A_2k_z^2 + A_3k_yk_z \pm \sqrt{\sum_{i=1}^{3} (B_i k_y + C_ik_z + D_ik_y^2 k_z + E_i k_yk_z^2)^2 }
\end{equation}
We use simulated annealing to fit this model to our surface band structure around the $\overline{Z}$ point. This gives the parameters shown in table I and II in SI appendix B. Note that our parameter space is very high dimensional, so we cannot hope to find the physically relevant parameters. This will be rectified in a future work. As we here are interested only in an effective model to compute the DOS, and as we have the full DFT results available, any model which reproduces the energy dispersion close to the crossing point is sufficient.


Extrema of the energy dispersion, $\nabla \varepsilon_{\mathbf{k}} = 0$, are called van Hove points and lead to singularities in the density of states (DOS): $g(E) = g_s \int \frac{d^{d-1}k}{(2 \pi)^{d}} \frac{1}{|\nabla \varepsilon_{\mathbf{k}}|_{\varepsilon_{\mathbf{k}} = E}}$ ($g_s$ denotes spin degeneracy). In two dimensions, $d = 2$, saddle points of the dispersion lead to a logarithmically divergent DOS. If this divergence occurs close to the Fermi energy, electronic correlations are strongly amplified, which can drive the system towards various electronically ordered states such as superconductivity, spin or charge density waves. 
In contrast, nodal points of the dispersion, where the Fermi surface shrinks to a point are characterized by a vanishing DOS.

\begin{figure}[t!]
\centering
\includegraphics[width=0.95\textwidth]{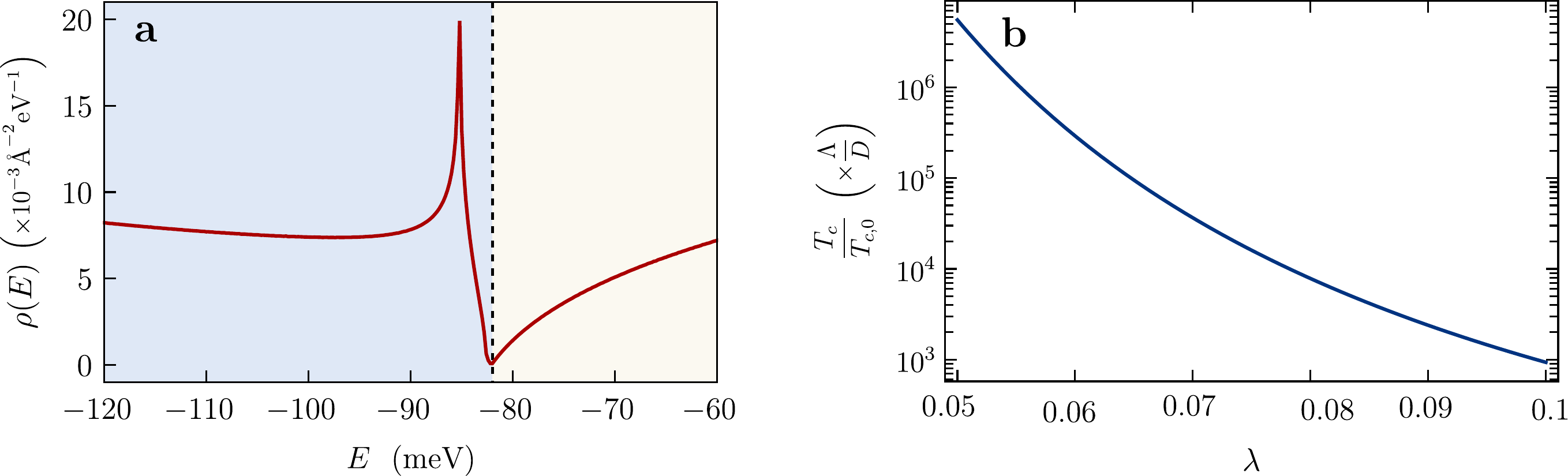}
\caption{
\textbf{a},  Density of states (DOS) of the $k\cdot p$ model describing the surface bands of RhBi$_2$ in the vicinity of the $\bar{\text{Z}}$ point (with coordinates $k_{\bar{\text{Z}},y}$ and $k_{\bar{\text{Z}},z}$) of the surface Brillouin zone. The DOS from states around $\bar{\text{Z}}$ exhibits a logarithmic divergence at $E = -85$~meV and vanishes at $E= -83$~meV. It is almost constant for energies $E < -85$~meV. Here, all energies are with respect to the Fermi energy at $E_F = 0$. {\bf b}, Enhancement of the superconducting transition temperature $T_c/T_{c,0}$ at the van Hove singularity as an function of the bare dimensionless superconducting coupling $\lambda=g\rho_0/2$ in the weak-coupling regime. Here, $T_{c,0}$ corresponds to the standard superconducting transition temperature obtained with a constant DOS at the Fermi level, while $T_c$ is the transition temperature due to a logarithmically divergent DOS at the Fermi level. $T_c$ is normalized by $D$, while $T_{c,0}$ is normalized by the cutoff $\Lambda$.
}
\label{fig:DOS}
\end{figure}

As described above, our ARPES measurements indicate the presence of both van Hove and nodal points in the surface bands of RhBi$_2$ that occur in close proximity to each other. To further underpin this observation, we here study in detail the properties of a low-energy $k\cdot p$ model (see Eq.~\eqref{eq:g}) that describes the band structure around the $\bar{\text{Z}}$ point in the surface BZ (sBZ). As shown in Fig.~ \ref{fig:4}a, the surface band structure exhibits a nodal point at the $\bar{\text{Z}}$ point, $(k_{\bar{\text{Z}},y}, k_{\bar{\text{Z}}, z}) = (0, 0.5)\frac{\pi}{a}$. At this point the singly degenerate lower and upper bands touch at an energy $E = -80$~meV below the Fermi energy $E_F = 0$. Both bands are anisotropic away from the nodal point due to the low symmetry of the material. This is particularly noticeable in the lower band, where the dispersion along the direction $k_y - k_{\bar{\text{Z}}, y} = -4.15 (k_z- k_{\bar{\text{Z}}, z})$ is almost flat and quadratic, while it is much steeper in the orthogonal direction (see Fig.S4 in the SI). Importantly, the lower band features two saddle points near $\bar{\text{Z}}$ at $(k_y,k_z) \approx (-0.042,0.510)\frac{\pi}{a}$ and $(0.042,0.490)\frac{\pi}{a}$. The saddle points reside at energy $E = -85$~meV, which is only $3$~meV below the energy of the nodal point. As shown in Fig.~\ref{fig:DOS}a, this gives rise to an intriguing density of states with a logarithmic divergence in close proximity to a zero. Finally, below the logarithmic singularity ($E < -85$ meV), the DOS is large and almost constant, corresponding to an effective mass of about $m^* \approx 0.4 m_e$. This is an average of the almost-flat character of the lower band along $k_{y}-k_{\bar{\text{Z}},y}=-4.15 (k_{z}-k_{\bar{\text{Z}},z})$ and the much steeper dispersion in the orthogonal direction.

To explore one of the possible consequences of the logarithmic divergence in the surface DOS, we calculate the enhancement of a hypothetical superconducting transition temperature $T_{c}$. For this purpose, we assume that a fully gapped $s$-wave superconducting state with a transition temperature $T_{c,0}$ is stabilized on the surface of RhBi$_2$ and we calculate the ratio $T_c/T_{c,0}$ as the system is tuned across the van Hove singularity. A discussion of the microscopic origin of superconductivity and possible other pairing channels is left for future work. Note that similar enhancements are expected in other interaction channels such as for density wave states.

Here, we focus on the weak coupling regime, where $T_{c}$ can be calculated from the standard BCS expression
\begin{equation}
1=g\int\limits_{-\Lambda}^{\Lambda} d\xi \, \frac{\rho(\xi)}{2\xi} \tanh\left(\frac{\xi}{2T_c}\right) \text{ ,}
\label{eq:gapBCS}
\end{equation}
\noindent where $g$ is a momentum-independent superconducting coupling, which is non-zero only inside an energy window of $2\Lambda$ around to the Fermi level. Note that in the case of phonon-mediated superconductivity, $\Lambda$ corresponds to the Debye frequency. Here, it has a more general meaning, as it denotes an energy cutoff for the effective superconducting coupling that could be mediated either by lattice vibrations (conventional pairing mechanism) or by the electrons themselves (unconventional pairing mechanism).

In the following, we set the Fermi level at the saddle points of the lower band, where the DOS diverges as $\rho(\xi)=\rho_0\log(D/\left|\xi\right|)$, with $\rho_0\approx 3.2\times 10^{-3}$~\AA$^{-2}$eV$^{-1}$ and $D\approx 16.5$~ meV. As a consequence, a large number of states in the vicinity of $\bar{\text{Z}}$ are available to pair with their time-reversed partners around $-\bar{\text{Z}}$ favoring a superconducting instability~\cite{hirsch1986enhanced_VHS_cal2}. Besides the apparent complication due to the logarithmic divergence of the DOS, Eq.(\ref{eq:gapBCS}) can be analytically solved to yield a superconducting transition temperature of
\begin{equation}
T_{c}\approx 1.13 D e^{-1/\sqrt{\lambda}} \text{ .}
\label{eq:TcLog}
\end{equation}

\noindent The quantity $\lambda=g\rho_0/2<1$ denotes a dimensionless superconducting coupling. Details about the derivation of Eq.~\eqref{eq:TcLog} can be found in SI Appendix~D Sec I. The hypothetical transition temperature $T_c$ is largely enhanced as compared to the standard BCS transition temperature $T_{c,0}\approx 1.13\Lambda e^{-1/\lambda}$, which is obtained from Eq.~\eqref{eq:gapBCS} for a DOS that can be approximated by a constant. This follows from the fact that Eq.~\eqref{eq:TcLog} depends on the dimensionless superconducting coupling $\lambda$ through $e^{-1/\sqrt{\lambda}}$ instead of $e^{-1/\lambda}$. Since $\lambda< 1$ in the weak coupling regime, this causes $T_{c}/T_{c,0} \gg 1$ as shown in Fig.\ref{fig:DOS}(b). Another (subleading) factor that contributes to the enhancement of $T_c$ is that Eq.~\eqref{eq:TcLog} is proportional to $D$ rather than $\Lambda$. Importantly $1/D$ gives the width of the DOS peak, i.e a larger $D$ corresponds to a sharper logarithmic divergence, and typically $D> \Lambda$~\cite{labbe1987superconductivity_VHS_cal1}.

Our results identify RhBi$_{2}$ as a very promising topological material that has a Dirac surface state with two saddle points close to E$_{F}$. DFT calculations and symmetry analysis show that triclinic RhBi$_{2}$ is a weak topological insulator having the simplest space group as reflected in its topological characterization. Based on the effective model, we show that the saddle point is related to a VHS. The proximity of the VHS to the Fermi level also provides opportunities to explore quantum many body instabilities to its inherent susceptibility to such instabilities. Our effective model and DFT calculations present guidance for understanding the saddle points and VHS around the Fermi level. 

Apart from the discussed implications, RhBi$_{2}$ also hosts potential for other future pursuits. For example, the material could host intriguing new physical phenomena and new topological phase transitions when stress or strain are applied.
On a related note, being a WTI in simplest form, RhBi$_{2}$ could also spark interest in making it a material platform to examine the role of defects. Indeed, growing a dislocation in different directions will ensure the binding of topologically protected modes  when the weak index vector is parallel to the defect's Burger vector \cite{defects1, defects2, defects3}. This pursuit is further underpinned by the unusual nature of the stacking vector (001;0) that is perpendicular to layering of the material and thus poses new relative orientations between the two, directly affecting the possibility of growing defects. 
\clearpage

\section*{References}

\bibliography{apssamp}

\clearpage
\section*{methods}

RhBi$_{2}$ crystals were grown using the high temperature solution growth method out of excess Bi\cite{Canfield2020RPP}. An initial concentration, Rh$_{20}$Bi$_{80}$, of elemental Rh and Bi was placed into a fritted alumina crucible \cite{Canfield2016PhilMag} and sealed in fused silica ampoule under a partial pressure of argon. The ampoule was then heated up to 900~$^{\circ}$C over 4 hours, held there for 3 hours and cooled down to 480~$^{\circ}$C over 200 hours. At this temperature the excess solution was separated from the RhBi$_{2}$ crystals using a centrifuge\cite{Canfield2020RPP}.

ARPES measurements were carried out using a laboratory based tunable VUV laser. The ARPES system consists of a Scienta DA30 electron analyzer, picosecond Ti:Sapphire oscillator and fourth-harmonic generator \cite{jiang2014tunable}. All data were collected with 6.7~eV photon energy. Angular resolution was set at $\sim$ 0.1$^{\circ}$ and 1$^{\circ}$, respectively, and the energy resolution was set at 2 meV. The size of the photon beam on the sample was $\sim 30\,\mu$m. Samples were cleaved \textit{in-situ} at a base pressure lower than 1$\times$ 10$^{-10}$ Torr, 40 K and kept at the cleaving temperature throughout the measurement.

Band structure with spin-orbit coupling (SOC) in density functional theory\cite{CAL_1_hohenberg1964inhomogeneous,CAL_2_kohn1965self} (DFT) have been calculated with PBE\cite{CAL_3_perdew1996generalized} exchange-correlation functional, a plane-wave basis set and projected augmented wave method\cite{CAL_4_blochl1994projector} as implemented in VAS\cite{CAL_5_kresse1996efficient,CAL_6_kresse1996efficiency}. The experimental lattice parameters in the triclinic unit cell are used. A Monkhorst-Pack\cite{CAL_7_monkhorst1976special} (7$\times$7$\times$7) \textit{k}-point mesh with a Gaussian smearing of 0.05 eV including the $\Gamma$ point and a kinetic energy cutoff of 229 eV have been used. To calculate topological properties, a tight-binding model based on maximally localized Wannier functions\cite{CAL_8_marzari1997maximally,CAL_9_souza2001maximally,CAL_10_marzari2012maximally} was constructed to reproduce closely the bulk band structure including SOC in the range of $E_{\textrm{F}}$ $\pm$ 1eV with Rh sd and Bi p orbitals. Then the spectral functions and Fermi surface of a semi-infinite RhBi$_{2}$ (100) surface were calculated with the surface Green's function methods\cite{CAL_11_lee1981simple,CAL_12_lee1981simple,CAL_13_sancho1984quick,CAL_14_sancho1985highly} as implemented in WannierTools\cite{CAL_15_wu2018wanniertools}.

\section*{Data availability}
Relevant data for the work are available at the Materials Data Facility Ref. XX

\section*{Acknowledgements}
We are grateful to Thomas Iadecola for useful discussions and Sergey L. Bud'ko for help with transport measurements. This work was supported by the U.S. Department of Energy, Office of Basic Energy Sciences, Division of Materials Sciences and Engineering. R.-J.~S.~acknowledges funding from Trinity college, the Marie Curie programme under EC Grant agreement No.~842901 and the Winton programme at the University of Cambridge. G.F.L acknowledges funding from the Aker Scholarship. The research (K.L., L.-L.W, B.K., T.V.T., N.H.J., P.P.O., P.C.C., and A.K.) was performed at Ames Laboratory. Ames Laboratory is operated for the U.S. Department of Energy by the Iowa State University under Contract No. DE-AC02-07CH11358. This work was also supported by the Center for Advancement of Topological Semimetals (B.K., T.V.T., N.H.J., P.P.O.), an Energy Frontier Research Center funded by the U.S. Department of Energy Office of Science, Office of Basic Energy Sciences, through the Ames Laboratory under its Contract No. DE-AC02-07CH11358.

\section*{Author contributions}
G.F.L, T.V.T., P.P.O and R.-J.S provided theoretical modelling and interpretation. B.K. and P.C.C. designed, grew and characterized the samples. L.-L.W. performed DFT calculations. K.L., N.H.J., B.S., and A.K. performed ARPES measurements and support. The manuscript was drafted by K.L., G.F.L., L.-L.W., B.K., T.V.T., N.H.J., P.P.O., R.-J.S., P.C.C. and A.K. All authors discussed and commented on the manuscript.

\section*{Competing interests}
The authors declare no competing interests.

\clearpage

\section*{ Supplementary information: Discovery of a weak topological insulating state and van Hove singularity in triclinic RhBi$_{2}$}

\section*{Appendix A: Crystal growth and characterization}
\label{appendixA}

 \begin{figure}[!htb]
    \includegraphics[scale=0.7]{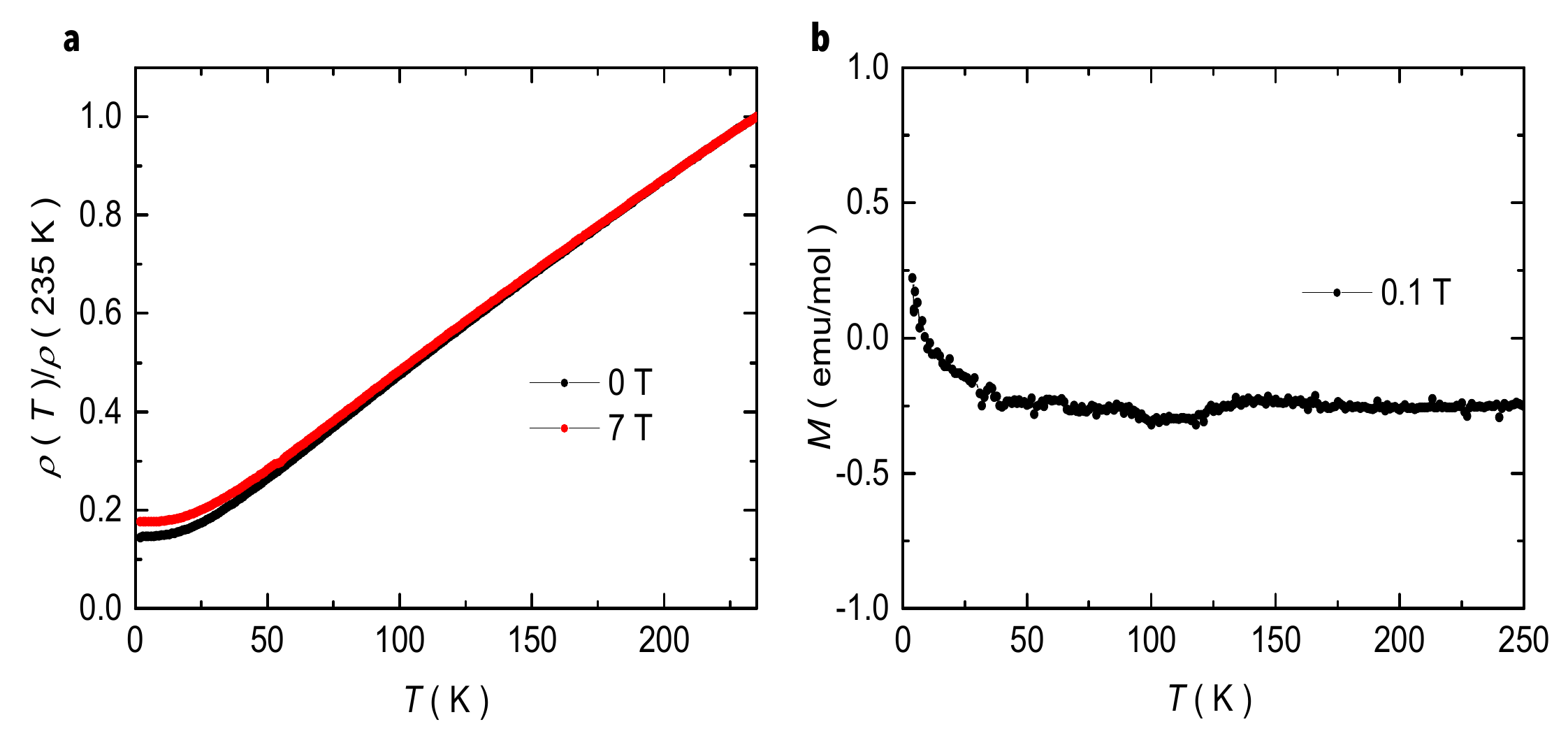}
    \caption{\textbf{Resistivity and magnetization data for RhBi$_2$.} {\bf a}, Resistivity data measured in zero field and 7 T. {\bf b}, magnetization measured as a function of temperature in 0.1 T field.}
  \label{}
\end{figure}

Resistivity as a function of temperature was measured with current in the $bc$-plane, along one of the crystallographic axis. $\rho$(T) shows clear metallic behavior with no features that would suggest any sort of phase transition in the measured temperature range. With an applied field of 7 T, $\rho$(T) do not show any significance changes, except slightly higher resistance values at low temperatures.  This very small magnetoresistance is not inconsistent with the relatively low RRR value ( $\sim$ 6) of the sample.   Magnetization as a function of temperature, M(T), was measured with field applied perpendicular to the $a$-axis, and shows nearly temperature independent behavior, with a small, negative value of M over most of the temperature range.  This fundamentally diamagnetic response is consistent with a very small density of states at the Fermi energy resulting in a larger core diamagnetic component.    

\section*{Appendix B: Details of the effective model}
\label{appendixB}
We aim to build an effective model for the observed surface state close to the $\overline{Z}$ point. As we know the form of the symmetries in the bulk, it would perhaps be most transparent to build a bulk model, which we then explicitly break to the relevant surface. In this section, we discuss this approach and explain why we instead opt to build a surface Hamiltonian directly. We also discuss the construction of this surface Hamiltonian.
\subsection{From bulk model to surface model}
A minimal bulk Hamiltonian requires four bands, due to the presence of doubly degenerate bulk Kramer's pairs. A general four-band model can be written in terms of $5$ $\Gamma$ matrices, and their corresponding commutators. However, we can restrict our model by imposing the bulk symmetries. The bulk has inversion and time-reversal symmetry. Combining these gives an operator $\mathcal{A} = \Theta \mathcal{P}$ which satisfies:
\begin{equation}
    \mathcal{A} H(\mathbf{k})\mathcal{A}^{-1} = H(\mathbf{k})
\end{equation}
We can choose our $\Gamma$-matrices to commute with the $\mathcal{A}$ operator\cite{fu2007IVS}. Requiring this gives that the commutators are all zero, so the effective model can be written using only the $5$ $\Gamma$-matrices. We can represent our symmetry operations as \cite{fu2007IVS}:
\begin{equation}
    \mathcal{P} = \sigma_x \otimes I, \quad \Theta = i(I\otimes s_y)\mathcal{K}
\end{equation}
Where $\mathcal{K}$ is complex conjugation. This gives the $5$ $\Gamma$-matrices commuting with $\mathcal{A}$ and satisfying the Clifford algebra as:
\begin{equation}
    \Gamma^{(1,2,3,4,5)} = \lbrace \sigma_x\otimes I, \sigma_y\otimes I, \sigma_z\otimes s_x,\sigma_z\otimes s_y, \sigma_z\otimes s_z \rbrace 
\end{equation}
Note that $\Gamma^1$ equals $\mathcal{P}$. We can then expand the Hamiltonian as:
\begin{equation}
    H(\mathbf{k}) = \epsilon I_{4\times 4}+M \Gamma^1 + \sum_{i=2}^5 A_i \Gamma_i = \epsilon I_{4\times 4}+M\Gamma^1 + S
\end{equation}
Now we note that:
\begin{equation}
    \Theta \Gamma^a \Theta^{-1} = \mathcal{P}\Gamma^a \mathcal{P}^{-1} =  \begin{cases}
      1 &\quad\text{if}\ a= 1\\
           -1 &\quad\text{if}\ a\neq  1
     \end{cases}
\end{equation}
From which it follows that $\epsilon$ and  $M$ are even in $\mathbf{k}\rightarrow -\mathbf{k}$, whereas the $\{A_i\}$ are odd. Furthermore, $S$ anticommutes with $\Gamma^1$, which which it follows that:
\begin{equation}\label{eq:parity_eigs}
\mathcal{P}H(k) \mathcal{P}^{-1} = \Gamma_1 H(k) \Gamma_1 = \epsilon I_{4\times 4}+M\Gamma_1 - S
\end{equation}
To get the surface states from this bulk Hamiltonian, we organize our Hamiltonian into a term $H_0$ which contains constants and $k_x$ dependent terms, and a perturbation $H_1$ which does not contain any $k_x$ dependence. As there is no preferred direction under the symmetry operations, $H_0$ contains all $5$ $\Gamma$-matrices, and is therefore not block diagonal. We can then in principle solve $H_0$ on a finite slab geometry, which  would give us four states, two of which are regular on the boundary. These can be interpreted as the edge states. From equation \ref{eq:parity_eigs}, it follows that the opposite parity states will be associated with an opposite sign for matrix $S$. Explicitly diagonalizing $H_0$ gives the eigenvectors:
\begin{equation}
   v_1 =  \begin{pmatrix}
    a\\
    b\\
    1\\
    0
     \end{pmatrix},    v_2 =\begin{pmatrix}
    c\\
    d\\
    0\\
    1
    \end{pmatrix},   v_3 = \begin{pmatrix}
    e\\
    f\\
    1\\
    0
    \end{pmatrix},     v_4 =\begin{pmatrix}
    g\\
    h\\
    0\\
    1
    \end{pmatrix}
\end{equation}
Where the states $(v_1,v_2)$ and $(v_3,v_4)$ have pairwise the same eigenvalues, and the coefficients are complicated functions of the free parameters. These are not block diagonal as $H_0$ is not block diagonal. We could now in principle investigate which of these four states remain regular at the boundary, given some boundary conditions, and then project the perturbation $H_1$ in this basis. Note, however, that the resulting expression will be rather convoluted, with many free parameters. As we are only interested in the surface dispersion to compute the DOS, we choose to instead build a model for the surface directly.

\subsection{Building the surface model directly}
The surface close to the TRIM $\overline{Z}$ exhibits time-reversal symmetry. Inversion symmetry is broken at the surface. However, there will be some residual effects of inversion symmetry, as otherwise $H_0$ should contain more terms, coming from the commutators of the $\Gamma$-matrices. Thus, inversion symmetry is broken down to some effective symmetry on the surface. The exact nature of this symmetry will depend on the orbital contents. Note in particular that if inversion acted purely as an effective $C_2$ symmetry, taking $(k_y, k_z)\rightarrow (-k_y, -k_z)$, then our expansion in equation \ref{eq:second_order_energy_kp} in the main text would be trivial. The exact nature of this effective symmetry can be elucidated by solving the full bulk Hamiltonian above. For our purposes, however, it suffices to ignore this residual symmetry, and only focus on the time-reversal symmetry. The residual inversion symmetry will only constrain the form of our effective Hamiltonian further, potentially resulting in less free parameters. Thus, we do not lose any physical solutions by only ignoring the residual inversion symmetry, though we may find an unphysical parameter combination. As we are only interested in the shape of the surface dispersion, this is of no concern.\\
\linebreak
The time-reversal symmetry $\Theta = i\sigma_y \mathcal{K}$ acts in spin space on the surface by constraining:
\begin{equation}
    H(\mathbf{k}) =(i\sigma_y)H^*(-\mathbf{k})(i\sigma_y)^{-1}
\end{equation}
We can write a two-band model for the surface states as:
\begin{equation}
    H(\mathbf{k}) = d(\mathbf{k})\cdot \sigma
\end{equation}
Hermiticity then requires that $d(\mathbf{k})$ should be real. Imposing time-reversal symmetry gives:
\begin{equation}
    d_0(\mathbf{k}) = d_0(-\mathbf{k}), \quad d_1(\mathbf{k}) = -d_1(-\mathbf{k}), \quad d_2(\mathbf{k}) = -d_2(-\mathbf{k}), \quad d_3(\mathbf{k}) = -d_3(-\mathbf{k})
\end{equation}
With energy given by:
\begin{equation}
    E(k) = d_0 \pm \sqrt{d_1^2 + d_2^2 + d_3^2}
\end{equation}
Expanding to second order then gives:
\begin{equation}
    H_2(k) = (E_0+A_1k_y^2 + A_2k_z^2 + A_3k_yk_z)I+(B_1k_y+C_1k_z)\sigma_x+(B_2k_y+C_2k_z)\sigma_y + (B_3k_y+C_3k_z)\sigma_z
\end{equation}
With energy
\begin{equation}\label{eq:second_order_energy_kp}
    E_{2\pm}(k) = E_0+A_1k_y^2 +A_2k_z^2 + A_3k_yk_z \pm \sqrt{\sum_{i=1}^3 (B_ik_y+C_ik_z)^2}
\end{equation}
If we want to account for strain, we should expand to third order which gives an effective model:
\begin{equation}
\begin{split}
    H_3(k)  = (E_0+A_1k_y^2 + A_2k_z^2 + A_3k_yk_z)I+(B_1k_y+C_1k_z+D_1k_y^2k_z + E_1k_z^2k_y)\sigma_x\\ +(B_2k_y+C_2k_z+D_2k_y^2k_z + E_2k_z^2 k_y)\sigma_y+(B_3k_y+C_3k_z+D_3k_y^2k_z + E_3k_z^2 k_y)\sigma_z
\end{split}
\end{equation}
With energy:
\begin{equation}\label{eq:third_order_energy_kp}
    E_{3\pm}(k) = E_0+A_1k_y^2 +A_2k_z^2 + A_3k_yk_z \pm \sqrt{\sum_{i=3}^3(B_ik_y+C_ik_z+D_ik_y^2k_z + E_ik_z^2 k_y)^2}
\end{equation}
Note that our energy is symmetric under $(k_y,k_z)\rightarrow (-k_y,-k_z)$ even though the Hamiltonian is not. This explains the observed effective $C_2$ symmetry in the FS.\\
\linebreak
The free parameters are fit using simulated annealing. This gives the result shown in table \ref{tab:second_order_param} and \ref{tab:third_order_param}.
\begin{table}[ht!]
    \centering
    \begin{tabular}{c c c c c c c c c}
         $A_1$& $A_2$ & $A_3$ & $B_1$ & $C_1$ & $B_2$ & $C_2$ & $B_3$ & $C_3$  \\
         \hline
         1.779& 3.827 & 0.886 & -0.248 & -1.225 & -0.138 & -0.252 & 0.185 & 0.273
    \end{tabular}
    \caption{Fit parameters for $E_{2\pm}$, see equation \ref{eq:second_order_energy_kp}}    
    \label{tab:second_order_param}
\end{table}
\begin{table}[ht!]
    \centering
    \begin{tabular}{c c c c c c c c c c c c c c c}
         $A_1$& $A_2$ & $A_3$ & $B_1$ & $C_1$ & $D_1$& $E_1$ & $B_2$ & $C_2$ & $D_2$&$E_2$& $B_3$ & $C_3$ & $D_3$ & $E_3$  \\
         \hline
         1.895& 8.051 & 2.701 & 0.145 & 0.021 & -41.264 & -150.345 & 0.082 & -0.154 & -2.524& -34.948 & 0.330 & 1.731 & 33.890 & 160.426
    \end{tabular}
    \caption{Fit parameters for $E_{3\pm}$, see equation \ref{eq:third_order_energy_kp}}    
    \label{tab:third_order_param}
\end{table}

\section*{Appendix C: DFT calculation of 2D cuts}
\label{appendixC}
\begin{figure}[!htb]
    \includegraphics[scale=0.6]{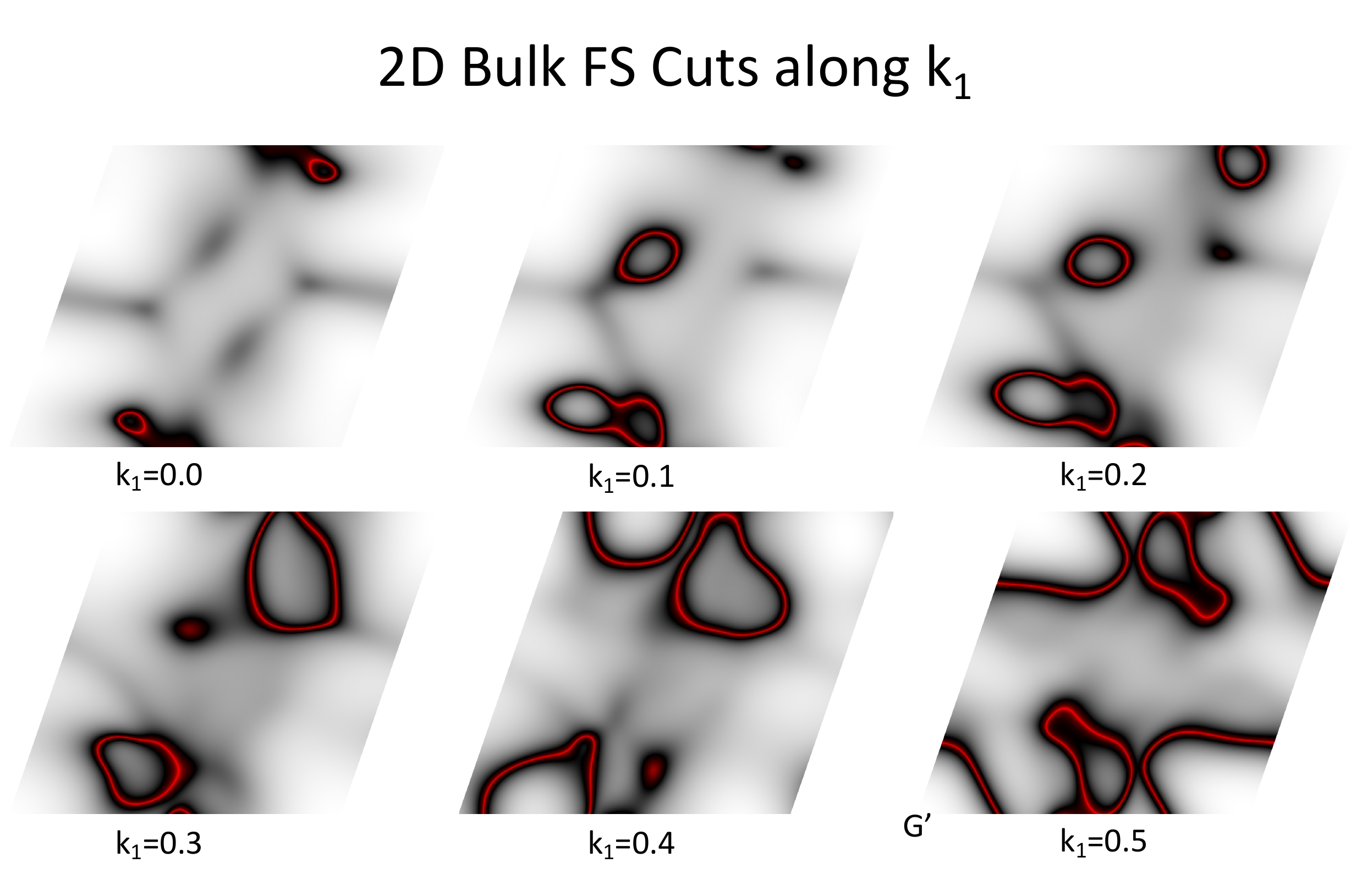}
    \caption{\textbf{FS projection along k$_{1}$ direction} 2D bulk Fermi surface cuts for RhBi2 along k$_{1}$=0.0, 0.1, 0.2, 0.3, 0.4 and 0.5 planes. The 2-fold rotation symmetry only appears on the k$_{1}$=0.0 and 0.5 planes.} 
 \label{fig:6}
\end{figure}

Although the triclinic structure of RhBi$_{2}$ only has inversion symmetry, $\mathbf{k} \rightarrow -\mathbf{k}$, when projected on the (100) surface, the inversion symmetry in 2D gives an effective twofold rotation symmetry. This can also be understood in the 2D cuts of the 3D FS as shown in Fig.\,\ref{fig:6}

\clearpage

\section*{Appendix D: band structure}
\label{appendixD}

A saddle point in the dispersion is characterized by a vanishing gradient, $\nabla \varepsilon_\mathbf{k} = 0$, and by opposite curvatures along orthogonal directions away from the extremum. Our ARPES measurements show evidence of the presence of a saddle point around the $\bar{\text{Z}}$ point in the surface Brillouin zone (sBZ) of RhBi$_2$. To precisely determine the position of the saddle point, we perform a close investigation of the band structure using an effective low energy $k \cdot p$ model valid close to $\bar{\text{Z}}$.

\begin{figure}[b!]
\centering
\includegraphics[width=0.6\textwidth]{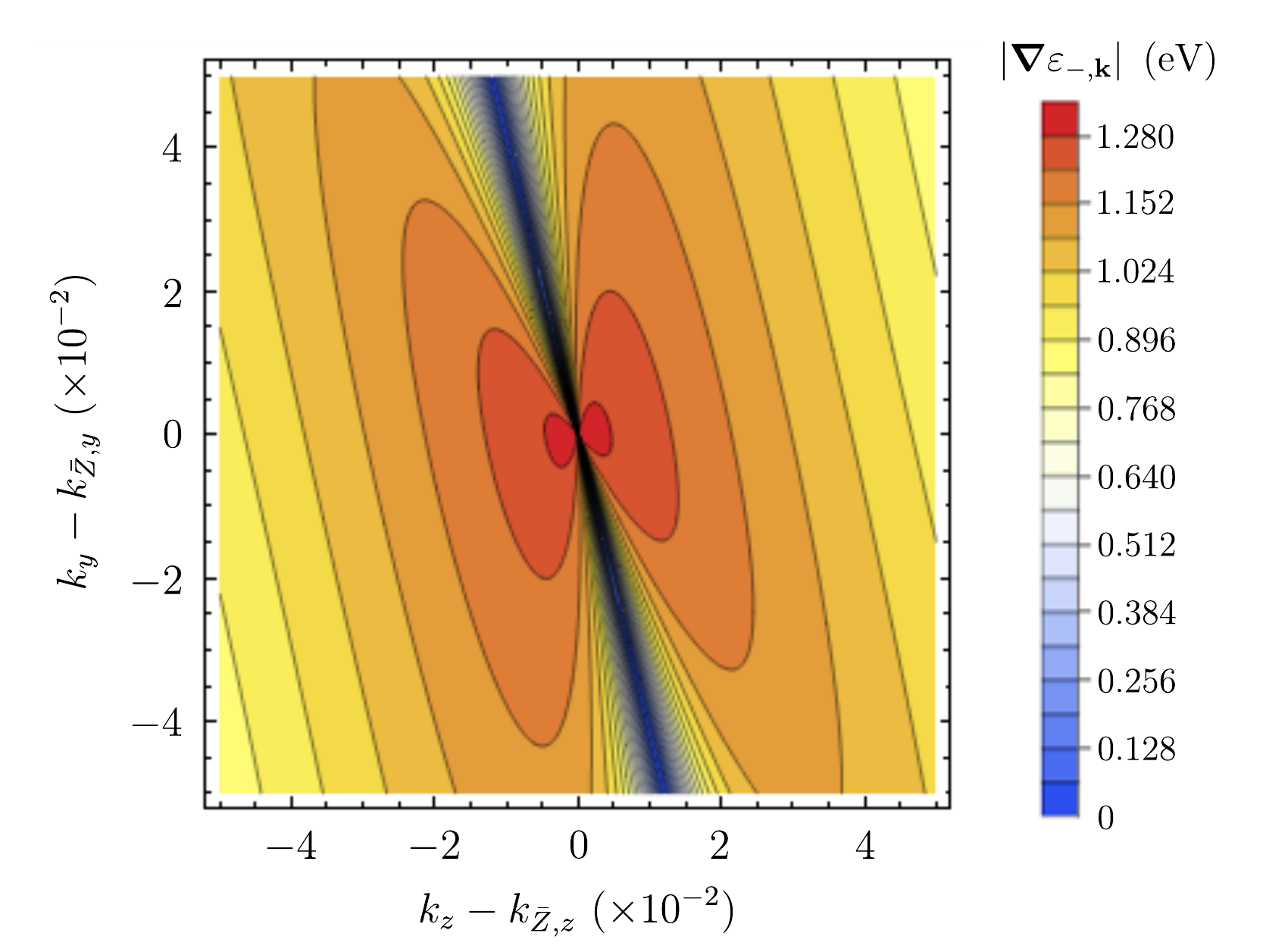}
\caption{\textbf{Gradient of the lowest energy band of the $k\cdot p$ model.} The direction where the norm of the gradient is minimum corresponds to the diagonal $k_y-k_{\bar{\text{Z}},y}=-4.15\left(k_z-k_{\bar{\text{Z}},z}\right)$. In particular, it vanishes at $(k_{y0},k_{z0})=\pm(-0.042,0.010)\frac{\pi}{a}$ with respect to $\bar{\text{Z}}$. These are saddle points of the lowest energy band. The momenta $k_y$ and $k_z$ are given in units of $\pi/a$. Accordingly, $|\nabla \varepsilon_{-,\mathbf{k}}|$ has dimensions of energy.}
\label{fig:gradient}
\end{figure}

\begin{figure}[t!]
\centering
\includegraphics[width=0.9\textwidth]{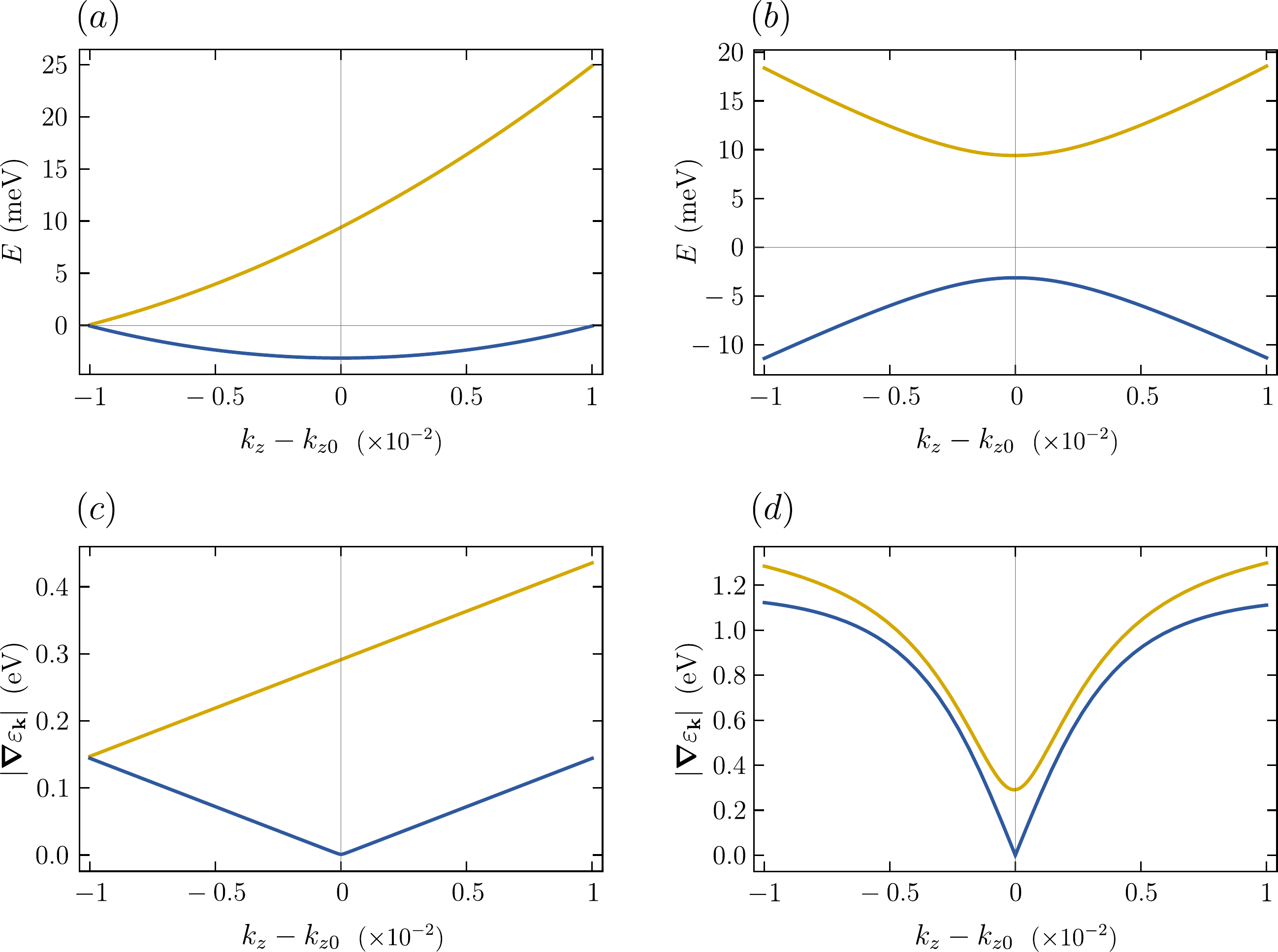}
\caption{Cuts of the band structure of the $k\cdot p$ model along the direction (a) $k_y-k_{\bar{\text{Z}},y}=-4.15\left(k_z-k_{\bar{\text{Z}},z}\right)$ and the orthogonal direction  (b) $k_y-k_{\bar{\text{Z}},y}=0.24\left(k_z-k_{\bar{\text{Z}},z}\right)$. Upper (lower) band is shown in yellow (blue). The dispersion is centered at the saddle point $\mathbf{k}_0=(-0.042,0.010)\frac{\pi}{a}$. Panels (c) and (d) show the norm of the gradient $|\nabla \varepsilon_\mathbf{k}|$ of the dispersion in {\bf a} and {\bf b}, respectively. The momenta $k_y$ and $k_z$ are given in units of $\pi/a$. Accordingly, $|\nabla \varepsilon_\mathbf{k}|$ has dimensions of energy.}
\label{fig:cuts}
\end{figure}

A first look into the band structure of the $k\cdot p$ model suggest the existence of two saddle points in the SBZ [see Fig.4 (a)]. Interestingly, as shown in Fig.\ref{fig:gradient}, the lower band $\varepsilon_{-,\mathbf{k}}$ is almost flat along the diagonal direction parametrized by
\begin{equation}
k_y=k_{\bar{\text{Z}},y}+\alpha (k_z-k_{\bar{\text{Z}},z}) \text{ ,}
\label{eq:diagonal}
\end{equation}
where $\alpha\approx 4.15$ and $(k_{\bar{\text{Z}},y},k_{\bar{\text{Z}},z})\approx (0, 0.5)\frac{\pi}{a}$, denotes the position the $\bar{\text{Z}}$ point. Zooming into the region around $\bar{\text{Z}}$ clearly shows two local minima along this flat direction at $ (k_y-k_{\bar{\text{Z}},y},k_z-k_{\bar{\text{Z}},z})=\pm(-0.042,0.010)\frac{\pi}{a}\equiv \pm\mathbf{k}_0$. These points, on the other hand, behave as local maxima of the dispersion along the direction orthogonal to Eq.~\eqref{eq:diagonal}. This behavior is evidenced in Figs.~\ref{fig:cuts}(a) and (b), respectively.

In fact, the gradient of the lower band $\nabla \varepsilon_{-,\mathbf{k}}$ vanishes at $\pm \mathbf{k}_0$ [see Figs.\ref{fig:cuts}(c) and (d)], which makes these points true saddle points of the lower band. They occur at energy $E = -83$ meV, where the peak in the surface density of states occurs [see Fig.4 (a)].

\section{Gap equation}
\label{app_sec:gap_equation}
In this section, we provide details on the analytic solution of the gap equation (5) in the main text in the presence of a logarithmically divergent density of states $\rho(\xi)=\rho_0\log(D/\left|\xi\right|)$. After the change of variables $y=\xi/T_c$, the gap equation can be rewritten as
\begin{equation}
1=g\rho_0\int\limits_{0}^{\Lambda/T_c} dy \log\left(\frac{D/T_c}{y}\right)\frac{1}{y}\text{tanh}\left(\frac{y}{2}\right) \,.
\end{equation}
Using properties of the logarithm, this integral can be further divided into two contributions,
\begin{equation}
\frac{1}{g\rho_0}=\log\left(\frac{D}{T_c}\right)\int\limits_{0}^{\Lambda/T_c}dy \frac{1}{y}\text{tanh}\left(\frac{y}{2}\right)-\int\limits_{0}^{\Lambda/T_c}dy\frac{\log(y)}{y}\text{tanh}\left(\frac{y}{2}\right) \text{ .}
\label{eq:gap2}
\end{equation}

\noindent The first integral in the right-hand-side of Eq.(\ref{eq:gap2}) is identical to the one encountered in the standard BCS gap equation when approximating the DoS by its (constant) value at the Fermi level:
\begin{equation}
\int\limits_{0}^{\Lambda/T_c}dy \frac{1}{y}\text{tanh}\left(\frac{y}{2}\right)=\log\left(\frac{\kappa \Lambda}{T_c}\right) \text{ ,}
\label{eq:term1}
\end{equation}

\noindent where $\kappa=2e^\gamma/\pi\approx 1.13$, and $\gamma$ is Euler's constant. For the second integral in Eq.(\ref{eq:gap2}), a simple integration by parts yields
\begin{equation}
\int\limits_{0}^{\Lambda/T_c}dy\frac{\log(y)}{y}\tanh\left(\frac{y}{2}\right)=\frac{1}{2}\log^2\left(\frac{\Lambda}{T_c}\right)-\frac{1}{4}\int\limits_{0}^{\Lambda/T_c} dy\log^2\left(y\right)\text{sech}^2\left(\frac{y}{2}\right) \text{ .}
\end{equation}
Since $\Lambda/T_c\gg 1$ and $\log^2\left(y\right)\text{sech}^2(y/2)$ remains finite as $y\rightarrow \infty$, we can approximate
\begin{equation}
\int\limits_{0}^{\Lambda/T_c} dy\log^2\left(y\right)\text{sech}^2\left(\frac{y}{2}\right)\approx \int\limits_{0}^{\infty} dy\log^2\left(y\right)\text{sech}^2\left(\frac{y}{2}\right)\approx \mathcal{C}\text{ ,}
\label{eq:term2}
\end{equation}

\noindent where $\mathcal{C}=2.669$ is a numerical constant. Substituting Eqs.(\ref{eq:term1}) and (\ref{eq:term2}) into Eq.(\ref{eq:gap2}), we obtain
\begin{equation}
\frac{1}{g\rho_0}=\log\left(\frac{D}{T_c}\right)\log\left(\frac{\kappa\Lambda}{T_c}\right)-\frac{1}{2}\log^2\left(\frac{\Lambda}{T_c}\right)+\frac{\mathcal{C}}{4} \text{ .}
\label{eq:res1}
\end{equation}

\noindent Since typically we have $D>\Lambda$ \cite{labbe1987superconductivity_VHS_cal1}, it is convenient to rewrite the logarithms in Eq.(\ref{eq:res1}) in terms of $D/T_c$ rather than $\Lambda/T_c$. This simplifies Eq.(\ref{eq:res1}) significantly, since the linear term in $\log(\kappa D/T_c)$ vanishes identically. We thus find
\begin{equation}
\frac{2}{g\rho_0}=\log^2\left(\frac{\kappa D}{T_c}\right)+\frac{\mathcal{C}}{2}-\log^2\left(\frac{\Lambda}{D}\right)-\log^2\left(\frac{1}{\kappa}\right) \text{ .}
\label{eq:res2}
\end{equation}

\noindent The strongest divergence comes from the term $\log^2(\kappa D/T_c)$, so that we can neglect the three last terms in the right-rand-side of Eq.(\ref{eq:res2}), which yields $T_c$ as given in Eq.(6) of the main text.

\end{document}